\documentclass[twocolumn,usenatbib]{mnras}

\usepackage{graphicx}
\usepackage{amsmath}

\def\pmb#1{\setbox0=\hbox{#1}%
    \kern-.025em\copy0\kern-\wd0
    \kern.05em\copy0\kern-\wd0
    \kern-.025em\raise.0433em\box0}

\def\ltsima{$\; \buildrel < \over \sim \;$}
\def\gtsima{$\; \buildrel > \over \sim \;$}
\def\simlt{\lower.5ex\hbox{\ltsima}}
\def\simgt{\lower.5ex\hbox{\gtsima}}
\def\p2Y{\;_2Y}
\def\m2Y{\;_{-2}Y}

\def\mk2{\mu {\rm K}^2}
\def\Planck{\it Planck\rm}

\def\LCDM{$\Lambda$CDM}

\newcommand{\Mpc}{\text{Mpc}} 
\newcommand{\Hunit}{~\text{km}~\text{s}^{-1} \Mpc^{-1}}

\def\pmb#1{\setbox0=\hbox{#1}%
     \kern-.025em\copy0\kern-\wd0
     \kern.05em\copy0\kern-\wd0
     \kern-.025em\raise.0433em\box0}

\begin{document}

\title[Evolving dark energy]{Evolving Dark Energy or Supernovae Systematics?}

\author[George Efstathiou]{George Efstathiou\\
 Kavli Institute for Cosmology Cambridge and 
Institute of Astronomy, Madingley Road, Cambridge, CB3 OHA.}

\maketitle

\begin{abstract}
  Recent results from the Dark Energy Spectroscopic Instrument (DESI)
  collaboration have been interpreted as evidence for
  evolving dark energy. However, this interpretation is strongly
  dependent on which Type Ia supernova (SN) sample is combined with
  DESI measurements of baryon acoustic oscillations (BAO) and
  observations of the cosmic microwave background (CMB) radiation. The
  strength of the evidence for evolving dark energy ranges from $\sim
  3.9 \sigma$ for the Dark Energy 5 year (DES5Y) SN sample to $\sim
  2.5 \sigma$ for the Pantheon+ sample.  The cosmology
  inferred from Pantheon+ sample alone is consistent with the \Planck\ \LCDM\
  model and shows no preference for evolving dark energy. In contrast, the
  the DES5Y SN sample favours evolving dark energy and is discrepant with
  the \Planck\ \LCDM\ model at about the $3\sigma$ level. Given
  these difference,  it is important to question whether they
  are caused by systematics in the SN compilations. A comparison of  
  SN common to both the DES5Y and Pantheon+ compilations shows evidence for an
  offset of $\sim 0.04$ mag. between low and high redshifts.  Systematics of
  this order can bring the DES5Y sample into  good agreement with the
  \Planck\ \LCDM\ cosmology and Pantheon+. I comment on a recent paper
  by the DES collaboration that rejects this possibility.
    
\end{abstract}


\section{Introduction}
\label{sec:Introduction}

The nature of dark energy remains an enigma following the discovery of
the accelerating Universe over 25 years ago \citep{Riess:1998,
  Perlmutter:1999}.  The simplest form of dark energy is a
cosmological constant with equation-of-state $w = p/\rho c^2 = -1$ \citep[see e.g. the review
  by][]{Weinberg:1989}. However, the quantum consequences of eternal
de-Sitter space lead to deep paradoxes suggesting that dark energy is
more complicated than a classical cosmological constant
\citep{Dyson:2002}.  One possibility is that our vaccum is metastable,
with a lifetime that is much longer than a Hubble time. From the
observational point of view, this proposal  is indistinguishable
from a cosmological constant.  An alternative possibility is that the
dark energy involves scalar fields and  is dynamical  \citep[for a
  comprehensive review see][]{Copeland:2006}. In such scenarios, the
equation-of-state can vary as a function of time,  offering an
intriguing challenge to observational cosmologists to detect, or
set limits,  on such a variation.

However, it is extremely difficult to detect a time variation in
$w$. In the case of a constant $w$, the observational constraints are
extremely tight. For example, the 2018 \Planck\ analysis
\citep[][]{Planck_Params_2018} found $w=-1.028 \pm 0.031$ by combining
\Planck\ CMB data, baryon acoustic oscillation (BAO) measurements and
the Pantheon SN catalogue \citep{Scolnic:2018}. The recent paper by
the DESI team \citep[][hereafter DESI24]{DESI:2024} combines
\Planck\ CMB data with BAO constraints from the first year DESI
observations  \citep{DESI_1:2024, DESI_2:2024} and the updated
Pantheon+ SN catalogue \citep{Scolnic:2022} to infer $w = - 0.997 \pm
0.025$. Despite the differences in the data, the results from these two analyses are
consistent and in excellent agreement with the \LCDM\ expectation of
$w=-1$.

The situation is very different if $w$ is allowed to vary with time. DESI24 parameterize the  evolution of
the equation-of-state with redshift $z$ as 
\begin{equation}
  w(z) = w_0 + w_a \left ( {z \over 1+z} \right ) ,  \label{equ:EoS1}
\end{equation}
\citep{Chevallier:2001, Linder:2003} introducing two additional parameters $w_0$ and $w_a$ compared to \LCDM. It is extremely difficult
to disentangle these two parameters. 
The CMB power spectra show a strong geometrical degeneracy between $w_a$ and $w_0$ along lines of constant angular diameter distance to the last scattering surface
\citep[see e.g. Fig.~30 of][]{Planck_Params_2018}  stratified by either $\Omega_m$ or $H_0$ (see e.g.
Fig.~ 6 of \cite{Efstathiou:2024}). The CMB power spectra cannot provide strong constraints on evolving dark energy because they
are  insensitive to departures from the \LCDM\ expansion history at  low redshift.  The BAO data cover redshifts  which are too high to  measure directly the effects of evolving dark energy favoured by the DES5Y SN data (which as we will see indicates deviations from  \LCDM\ the background expansion rate at $z \simlt 0.2$). In fact,
BAO measurements over the wide redshift range $0.3 - 2.5$, including the new DESI BAO  measurements,  are consistent with the \LCDM\ cosmology \citep[e.g.][]{Alam:2017, Planck_Params_2018, Alam:2021, DESI:2024}. BAO measurements therefore affect inferences on low redshift variations in $w$ indirectly by imposing a constraint on  $\Omega_m$. Direct inference on variations in $w$ at low redshift is therefore strongly reliant on the magnitude-redshift relation of Type Ia SN.

Type Ia SN are inherently complex objects. However,  heroic efforts over
many years have led to 'standardized' SN compilations that have been
used extensively  to test cosmology. The 2018 \Planck\ papers adopted the Pantheon 
SN sample \citep{Scolnic:2018} superceeding the earlier SNLS \citep{Conley:2011},
Union2.1 \citep{Suzuki:2012} and JLA \citep{Betoule:2014} compilations. DESI24 used three
SN catalogues in their analysis of cosmological parameters: the Pantheon+ sample of
\cite{Scolnic:2022}, the Union3 compilation of \cite{Rubin:2023} and the DES5Y sample
\citep{Sanchez:2024, Vincenzi:2024, DES5Y:2024}. DESI24 report evidence for a time
varying equation-of-state with $w_0>-1$ and $w_a <0$, but with a statistical significance that
is strongly dependent on which SN sample was chosen to be  combined  with  \Planck\ CMB and DESI
BAO measurements. Specifically, DESI24  find $\sim 2.5\sigma$, $3.5\sigma$ and $3.9 \sigma$ evidence
for dynamical dark energy using the Pantheon+, Union3 and DES5Y SN samples respectively.  These numbers are reduced slightly to $\sim 2.2 \sigma$, $3.1\sigma$ and $3.5\sigma$  respectively using a combined
DESI+SDSS BAO compilation as described in Section 3.3 of DESI24. The Pantheon+ sample (as we will show below) is consistent with the \Planck\ \LCDM\ cosmology and so the DESI24 results for this SN compilation could plausibly be interpreted as a statistical fluctuation. However, the high
statistical significance levels for the Union3 and DES5Y SN samples are unlikely to be statistical fluctuations and therefore deserve scrutiny.

In this paper, I investigate the statistical consistency of the
Pantheon+ and DES5Y SN samples. These two catalogues show the largest differences in
the DESI24 analysis. The Union3 compilation for individual SN is not
yet in the public domain and would, in any case, require a very
different type of analysis to that presented here. The Pantheon+/DES5Y
comparison is particularly simple because the DES5Y compilation has
only a small number of SN at low redshifts with a high degree of
overlap with low redshift SN in the Pantheon+ sample.

Section \ref{sec:SNsamples} describes the Pantheon+ and DES5Y SN
samples and summarizes cosmological constraints on the parameters of
the \LCDM\ cosmology and  the $w_0$, $w_a$
parameterisation of Equ.~\ref{equ:EoS1} (which we denote as $w_0w_a$CDM). We cross-correlate
Pantheon+ with DES5Y in Sect. \ref{sec:PanvDES}, finding evidence for a systematic offset of $\sim 0.04$ mag
between low and high redshifts for SN common to both catalogues.\footnote{We note that in a recent paper \cite{Dhawan:2024} discuss various potential sources of
redshift dependent systematic errors in SN photometry.} We then discuss the cosmological implications of
systematic offsets of this order. In Sect.~\ref{sec:DESreply} I comment on the recent paper by the DES collaboration
\citep{DESreply:2025} which forcefully  stands by the validity of the DES5Y SN results.
Sect. \ref{sec:conclusions} summarizes our conclusions. Appendix A discusses fits to the DESI BAO and in combination with the Pantheon+ and DES5Y SN compilations.

\section{The Pantheon+ and DES5Y SN samples }
\label{sec:SNsamples}

The construction of the Pantheon+ catalogue is complex as summarized in the flow diagram in Fig. 1 of
\cite{Brout:2022}. The catalogue `standardizes' observations from 20 supernova compilations spanning a wide redshift range from the local SN used in the SH0ES distance scale project \citep{Riess:2022} to a maximum redshift of $2.261$. The cross-calibration of the photometric systems and corrections for biases, which involve forward modelling of the SN populations, are described in \cite{Brout:2021, Popovic:2021,  Brout:2022b, Popovic:2023}. The Pantheon+
catalogue contains $203$ spectroscopically confirmed Type Ia SN from the DES year 3 sample \citep{Brout:2019, Smith:2020}.

\begin{figure}
	\centering
  \includegraphics[width=85mm, angle=0]{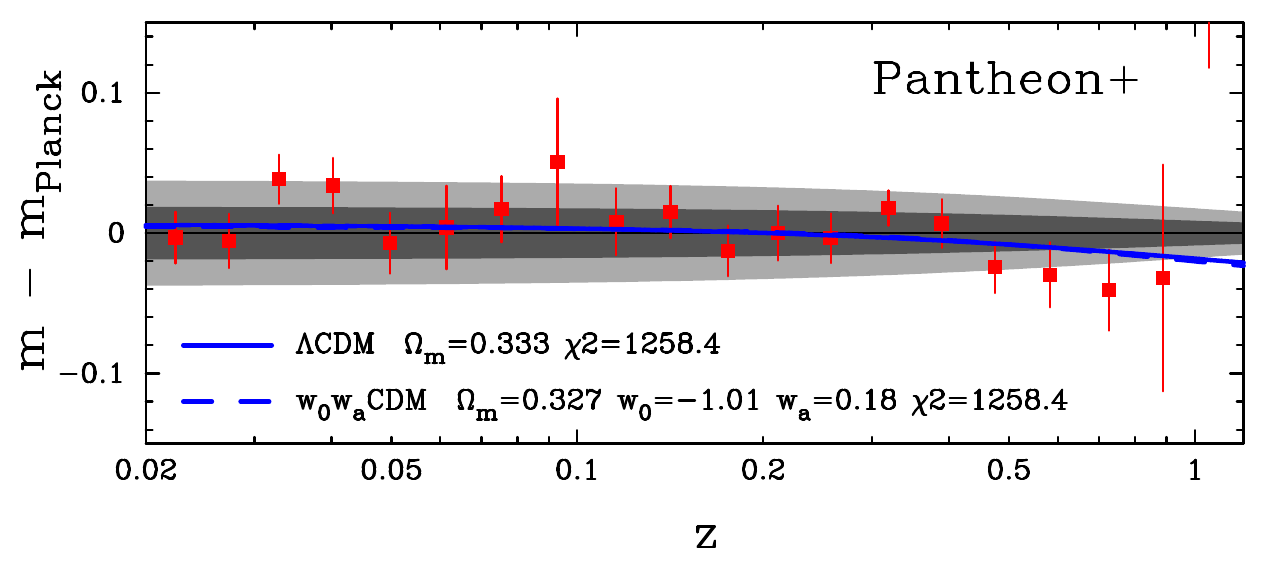} \\  \includegraphics[width=85mm, angle=0]{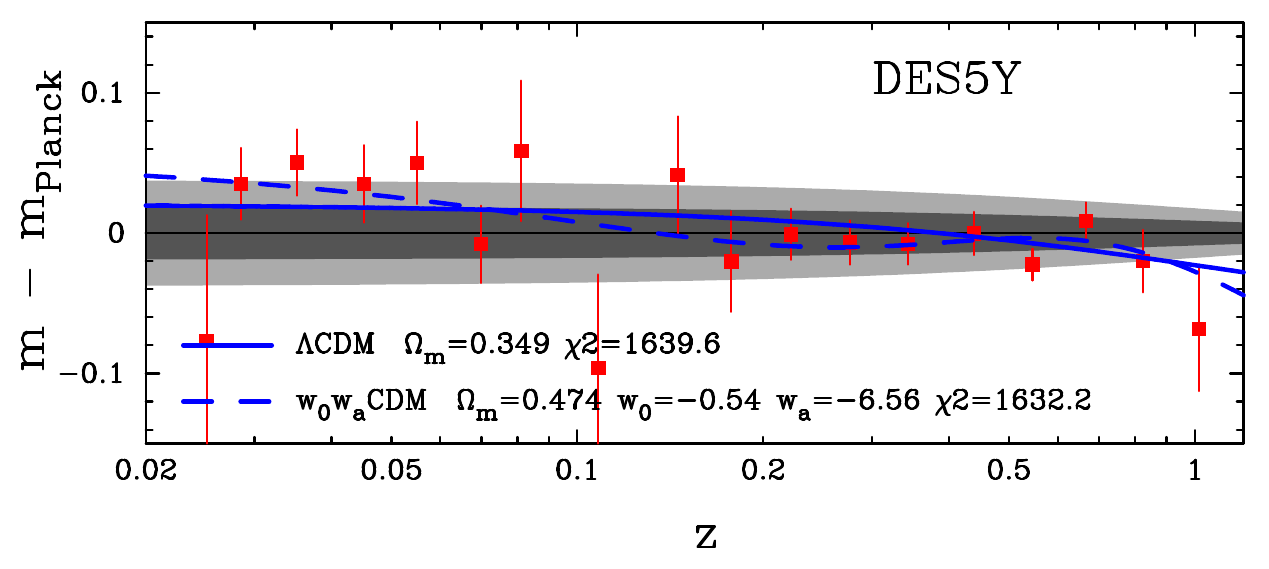} 
  \caption{Residuals of the magnitude redshift relation relative to that expected in the best fit Planck TTTEEE
    \LCDM\ cosmology. The grey bands show the \Planck\ $\pm 1\sigma$ and $\pm 2 \sigma$ error ranges. The red points show the
    maximum likelihood residuals for the SN data averaged  in logarithmically space bins in (Hubble distance) redshift. The errors are computed from the full covariance matrices for the two SN catalogues. The upper panel shows the fits for the Pantheon+ catalogue and the lower panel for the DES5Y catalogue. The solid blue lines show the best fit \LCDM\ cosmology
    and the dashed lines show the best fit $w_0w_a$CDM cosmology. The best fit parameters are listed in the Figure.
    Our fits retained $1417$ and $1754$ SN in the Pantheon+ and DES5Y samples respectively. }

	\label{fig:fits}

\end{figure}

\begin{figure*}
  \center
  \includegraphics[width=55mm, angle=0]{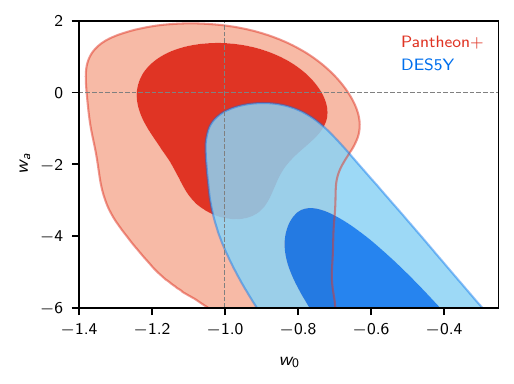}
  \includegraphics[width=55mm, angle=0]{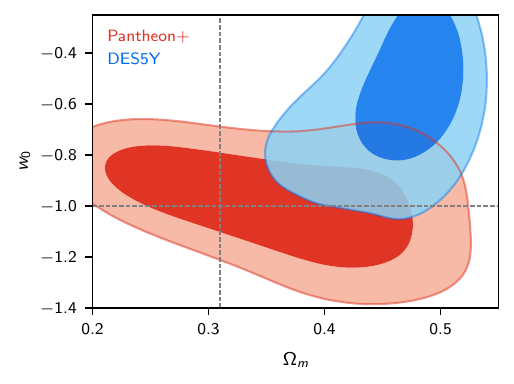}
  \includegraphics[width=55mm, angle=0]{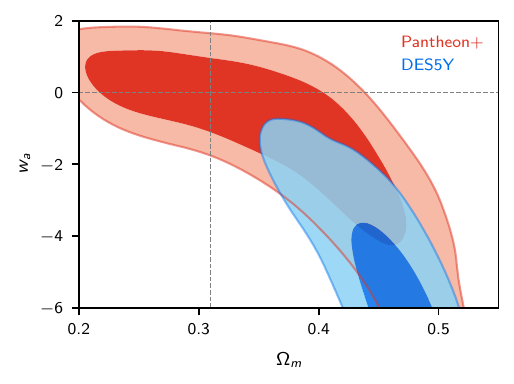}   
  \caption{68\% and 95\% contours for the marginalized posterior parameters in the $w_0-w_a$, $w_0-\Omega_m$ and
    $w_a-\Omega_m$ planes for the Pantheon+  (red) and DES5Y (blue) SN compilations. The dashed lines show the values for
  the best fit \Planck\ \LCDM\ cosmology.}

	\label{fig:PanDES}

\end{figure*}

The DES5Y SN sample consists of $1635$ DES SN spanning the redshift range $0.0596$ to $1.12$ together with $194$ SN at redshifts $z<0.1$ from a
small number of more recent (post 2009) SN surveys. Historical low redshift SN samples are not included to simplify the cross-calibration analysis.
As with Pantheon+, the construction of the DES5Y sample is complicated and is most succinctly summarized in \cite{Vincenzi:2024}. Several  aspects
of the DES5Y analysis are similar to those in Pantheon+. However,  since most of the DES supernovae lack spectroscopic classification, the DES team apply the Bayesian classifier SuperNNova \cite[see][and references therein]{Kessler:2017, Moller:2020}
to compute a probability that a SN belongs to the Type Ia population. The uncertainty in photometric classification is propagated into the
covariance matrix. The catalogue lists the distance moduli, $\mu_{\rm DES}$,  of the SN assuming a Hubble constant of $70 \Hunit$. To convert to the magnitude system of
Pantheon+ I apply the correction:
\begin{equation}
  m_{\rm DES} = \mu_{\rm DES} - 19.33,   \label{equ:calib}
 \end{equation}
based on my fits of the \Planck\ best fit \LCDM\ cosmology to the Pantheon+ sample to determine the SN peak absolute magnitude.

Figure \ref{fig:fits} shows the residuals of the magnitude-redshift relations relative to the best fit \Planck\ TTTEEE \LCDM\ fit\footnote{Computed from the 12.5cln chains of \cite{Efstathiou:2021}.}. Here I compute the maximum likelihood value of the SN apparent magnitude $\overline m_k$ in $N_k$ logarithmically spaced bins in redshift together with the diagonal errors computed from the full covariance matrices $C_{ij}$ of the respective SN samples.  Thus
\begin{subequations}
\begin{equation}
\overline m_k =   \sum_{i \subset k}\sum_{j \subset k } m_i C^{-1}_{ij} /\sum_{i \subset k}\sum_{j \subset k }C^{-1}_{ij} , \label{equ:bin1}
\end{equation}
with covariance matrix
\begin{eqnarray}
  \hspace {-0.12 truein} \langle \delta \overline m_k \delta \overline m_{k^\prime} \rangle \hspace {-0.12 truein} &=&  \left.  \hspace {-0.12 truein}  \sum_{i \subset k}\sum_{j \subset k }\sum_{p \subset k^\prime}\sum_{q \subset k^\prime } C_{ip} C^{-1}_{ij}C^{-1}_{pq} \middle / \right.
    \qquad \qquad \nonumber  \\
 \hspace {-0.12 truein}& &\ \hspace {-0.12 truein} \left[\sum_{i \subset k}\sum_{j \subset k }C^{-1}_{ij}\right]\left[\sum_{p \subset k^\prime}\sum_{q \subset k^\prime }C^{-1}_{pq}\right] ,  \label{equ:bin2}
\end{eqnarray}
\end{subequations}
where the sums extend over all supernovae in bins $k$ and $k^\prime$.
Note that the binning is for plotting purposes only. All of the fits discussed in this paper have been done on the unbinned data, though fits to the binned Hubble diagram with the covariance matrix of Eq.~\ref{equ:bin2} give almost identical results.  We use SN with `Hubble distance' redshifts, $z_{\rm HD}$ (i.e. corrected to the CMB frame and for peculiar velocities) and for both samples we use SN in the redshift range $0.02 \le z_{\rm HD} \le 1.2$. In addition, we remove SN from DES5Y with magnitude errors $\delta m >1$, since these
have a low probability of being Type Ia SN based on the DES photometric data. This leaves $1417$ entries
in the Pantheon+ sample (out of a total of $1701$ entries; note that in a few cases, Pantheon+ lists separately two or more observation of the same SN) and $1754$ SN in DES5Y (out of a total of 1829). For reference, the $\chi^2$ values for the \Planck\ \LCDM\ cosmology ($\Omega_m = 0.3135 \pm 0.0081$) are $1259.6$ for Pantheon+ and
$1644.68$ for DES5Y.

\smallskip

\noindent
(i) Pantheon+: The \LCDM\ best fit gives $\Omega_m = 0.333 \pm 0.018$,
slightly higher than the value of $\Omega_m$ for best fit
\Planck\ \LCDM\ cosmology and in agreement with the results of
\cite{Brout:2022}. The $\chi^2$ value for this fit improves by only
$\Delta \chi^2 = 1.2$, thus the Pantheon+ sample is consistent with
the \Planck\ \LCDM\ cosmology. There is no improvement in $\chi^2$
compared to \LCDM\ if $w_0$ and $w_a$ are allowed to vary. The
Pantheon+ SN data therefore offer no support for models with a time
varying equation-of-state.

\smallskip

\noindent
(ii) DES5Y: The \LCDM\ best fit gives $\Omega_m = 0.354\pm 0.017 $
which is $\sim 2.2\sigma$ higher than the best fit \Planck\ \LCDM\ value with an
improvement in $\chi^2$ of $\Delta \chi^2 = 5.1$.  However, allowing
$w_0$ and $w_a$ to vary leads to a substantial reduction in $\chi^2$,
and visually to a much better fit to the magnitude residuals. By adding $w_0$ and $w_a$, the best fit is able to track the inflection at the transition 
between the low redshift SN and the  DES SN (with smaller error bars on the binned magnitude residuals) at $z> 0.2$.
The DES5Y compilation clearly prefers evolving dark energy, with $\Delta
\chi^2 = 12.5$ relative to the best fit \Planck\ \LCDM\ cosmology and $\Delta
\chi^2 = 7.4$ relative to  the best fit DES5Y \LCDM\ cosmology.  Thus, the
\Planck\ \LCDM\ cosmology is strongly disfavoured by the DES5Y data
(at roughly $3\sigma$), and evolving dark energy is preferred over
the best fit DES5Y \LCDM\ cosmology at $\sim 2 \sigma$.

The tension between the Pantheon+ and DES5Y datasets is clear from the
parameter constraints plotted in Fig.~\ref{fig:PanDES}.  To produce
Fig.~\ref{fig:PanDES},  I applied uniform priors on the parameters
$\Omega_m$, $w_0$ and $w_a$ as follows: For Pantheon+, $0.2 \le
\Omega_m \le 0.6$, $-1.4 \le w_0 \le 0.4$, $-4 \le w_a \le -0.4$; For
DES5Y, $0.01 \le \Omega_m \le 0.81$, $-2 \le w_0 \le 1$, $-10 \le w_a
\le 1$. I applied a \Planck\ \LCDM\ Gaussian prior on $H_0$ with
$67.44 \pm 0.58 \Hunit$ and treated the SN absolute magnitude $M$ as a
derived parameter. Since we are not combining the SN data with other
cosmological data, the prior on $H_0$ has no impact on the results
plotted in Fig.~\ref{fig:PanDES}. Chains were computed with the
{\tt MULTINEST} nested sampler \citep{Feroz:2009, Feroz:2011}\footnote{
The {\tt MULTINEST} chains used to produce Fig.~\ref{fig:PanDES} agree well with the emcee \citep{emcee:2013} chains available on the DES web site:
\noindent
    {\tt https://github.com/des-science/DES-SN5YR/tree/main}
    \noindent
        {\tt 5$\_$COSMOLOGY/chains).} }.  . The best fit
\Planck\ \LCDM\ parameters are shown by the dotted lines which
intersect within the $1\sigma$ contours for the Pantheon+ sample.  In
contrast, the DES5Y sample wants high values of $\Omega_m$, $w_0> -1$
and and low values of $w_a \simlt -2$, favouring dynamical dark
energy. This, of course,
is simply a consequence of the redshift dependence of the DES5Y magnitude-redshift relation shown in Fig.~\ref{fig:fits}.  The favouring of high values of $\Omega_m$ is especially worrying since many different cosmological probes, including the CMB, BAO, galaxy power spectrum full shape analysis \citep[e.g.][]{Philcox:2022, dAmico:2024},  cluster baryon  fractions \citep{Mantz:2022} give
concordant values of  $\Omega_m \approx  0.3$ to within a few percent, 
in strong disagreement with the values $\Omega_m \simgt 0.4$ favoured by DES5Y. The results of combining the DES5Y and Pan+
samples with the DESI Y1 BAO measurements are summarized in Appendix \ref{sec:appendix}.

\section{Comparison of Pantheon+ and DES5Y SN}
\label{sec:PanvDES}

Figure \ref{fig:PanDES} poses the question of whether the Pantheon+
and DES5Y SN samples are statistically compatible. Instead of using
the consistency of the  cosmological parameters shown in Fig.~\ref{fig:PanDES}, 
 which are relatively poorly constrained, to tackle this question it is informative to
compare directly the magnitudes of identical SN reported in the two
catalogues. If the respective SN compilations  are accurate at the $\sim 0.01$ mag level (necessary for unbiased cosmology) 
it should be possible to understand any differences seen in such a comparison to this accuracy.

\begin{figure}
  \center
  \includegraphics[width=80mm, angle=0]{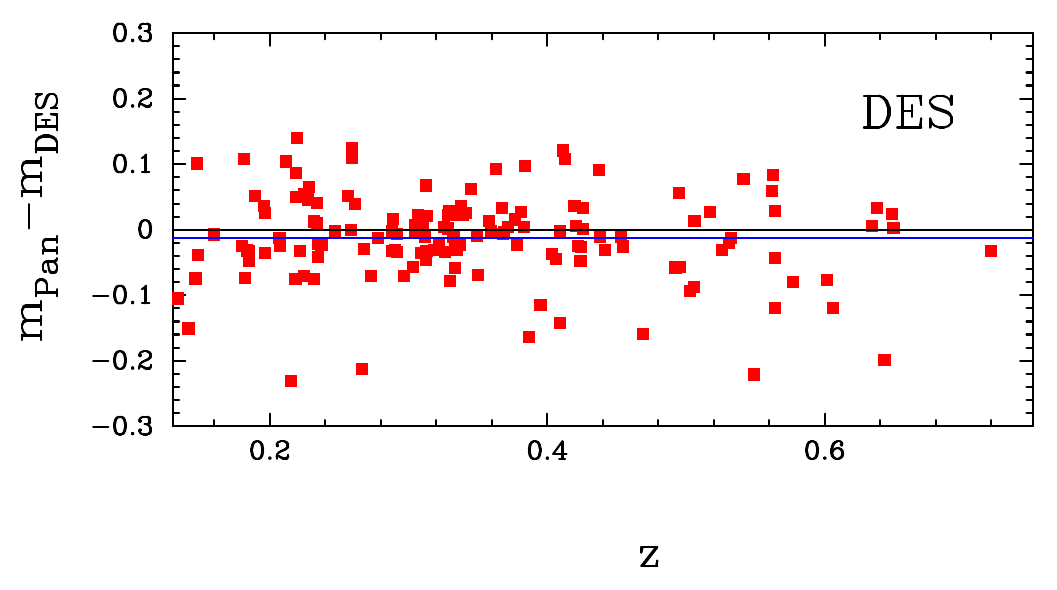} \\
  \includegraphics[width=82mm, angle=0]{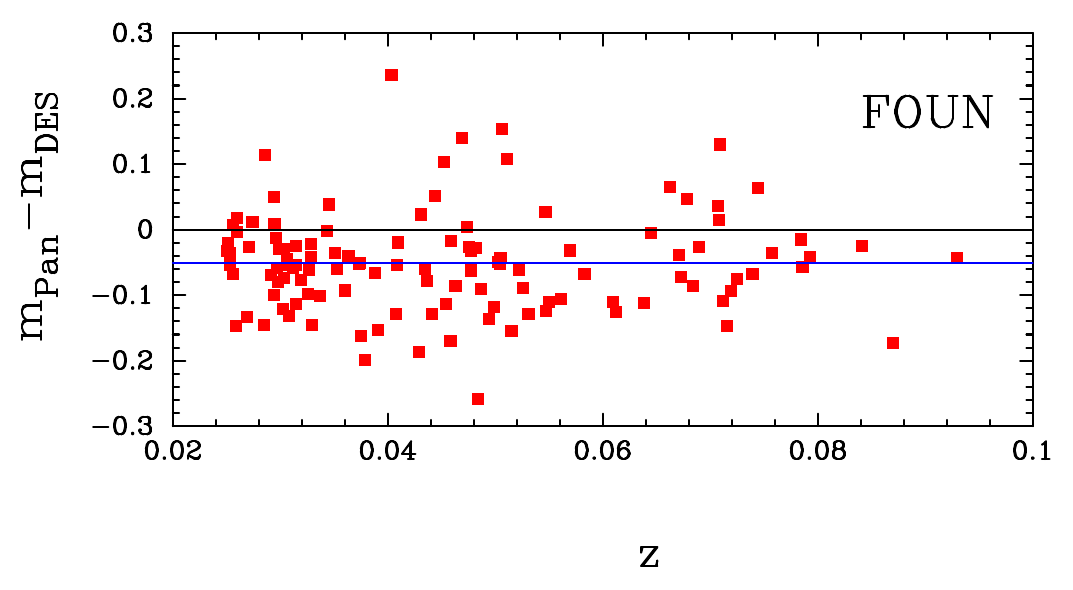}   
	\caption{ The upper panel shows the magnitude differences  as a function of redshift for DES SN  common to both the Pantheon+ and DES5Y catalogues. The blue line shows the mean value giving each SN equal weight. The lower panel shows the magnitude differences for SN from the low redshift Foundation Supernova Survey \citep{Foley:2018}.  }

	\label{fig:PanDESmags}

\end{figure}

The Pantheon+ compilation includes 203 DES Y3 spectroscopically confirmed Type Ia
SN. $145$ of these SN are contained in the DESY5 compilation. The
magnitude differences for these SN are plotted against Hubble distance
redshift in the upper panel of Fig.~\ref{fig:PanDESmags}. The offset
applied to the DESY5 distance moduli (Equ.~\ref{equ:calib}) has been
chosen to give a SN absolute magnitude that matches the best fit
for the Pantheon+ SN assuming the \Planck\ \LCDM\ value
of $H_0$. It is important to emphasise that the  main aim of this paper is
assess the impact of differential magnitude differences between low and high redshifts
{\it within a SN compilation} and therefore the value  of the offset is irrelevant.
With the choice of Equ.~\ref{equ:calib}
the mean offset is for DES SN common to both catalogues us close
to zero,  $\langle m_{\rm Pan} - m_{\rm DES} \rangle = -0.012 \ {\rm mag}$.
The dispersion around the mean is $0.067$ mag., substantially
smaller than the errors on the magnitudes listed in the Pantheon+ and
DES5Y catalogues. The  scatter seen in this plot 
may  reflect difference in the light curve fitters used in the two samples, SALT2 \citep{Guy:2007}
for Pantheon+ and SALT3 \citep{Kenworthy:2021} for DES5Y.
In addition, the  two catalogues 
have different bias corrections as described in \cite{Brout:2022b}, 
\cite{Vincenzi:2024} and in Sect. \ref{sec:DESreply} below.

\begin{figure}
  \center
  \includegraphics[width=85mm, angle=0]{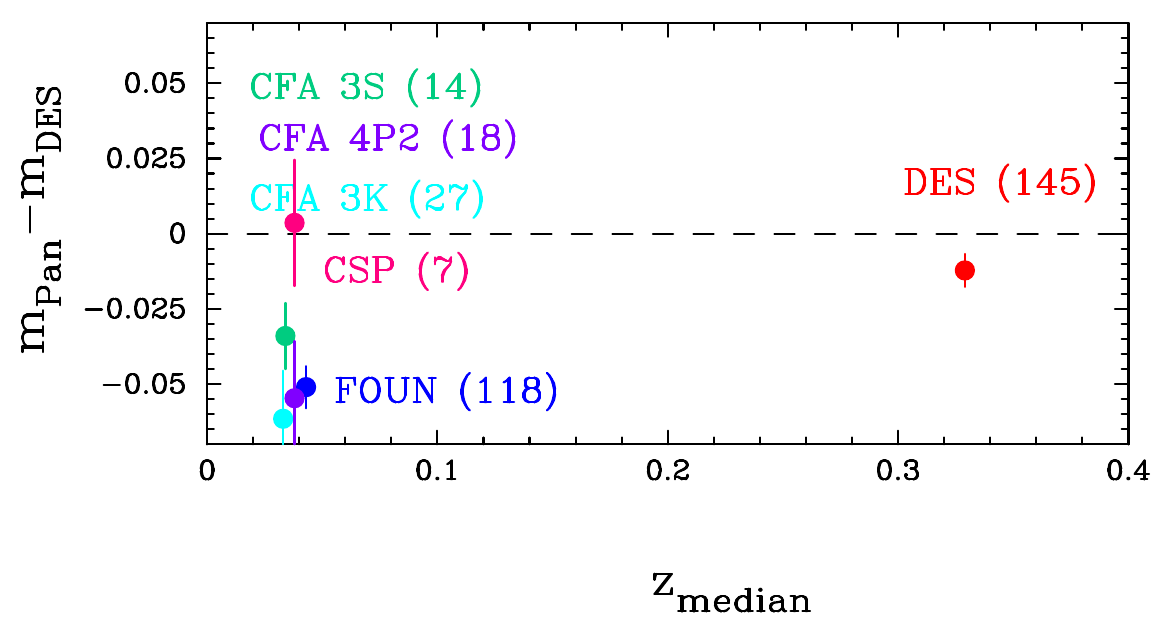} 
  \caption{The mean magnitude offset for SN common to Pantheon+ and DES5Y grouped by source catalogue. The numbers
    in brackets give the the number of SN common to both surveys. There are 3 SN in common in the CfA4P3 sample
  with a mean magnitude difference of  of $0.028 \pm 0.087$ mag. which is not plotted because the error is so large (and unreliable). }

	\label   {fig:PanDES_offset}

\end{figure}

The DES5Y compilation uses only a small number of SN samples to probe the low redshift end ($z \simlt 0.1$) of the magnitude-redshift relation.  The largest of these is the Pan-STARRS Foundation Supernova Survey \citep{Foley:2018}.
150 SN from the Foundation Survey are included in the Pantheon+ catalogue and   $118$ of these are also included in the DES5Y
catalogue. The magnitude differences are plotted in the lower panel of Fig.~\ref{fig:PanDESmags}. This shows a systematic offset
between the Pantheon+ and DES5Y magnitudes of $\langle m_{\rm Pan} - m_{\rm DES} \rangle = -0.051 \pm 0.007 \ {\rm mag}$ (giving each SN equal weight). In addition, the DES5Y catalogue includes 68 SN from the Center for Astrophysics (CfA)  SN surveys
\citep{Hicken:2009, Hicken:2012} and 8 SN from the Carnegie Supernova Project \citep[CSP, ][]{Krisciunas:2017}, 7 of which are included in Pantheon+. Figure \ref{fig:PanDES_offset} plots the Pantheon+/DES5Y offsets for these surveys. Most of the low redshift surveys are offset by $\sim 0.04 -  0.05$ mag. relative to the more distant DES supernovae.
Table~\ref{tab:magfits} lists the median redshift and mean magnitude offset for each sample.
Offsets of this order can
easily explain the  differences between the Pantheon+ and DES5Y magnitude-redshift relations seen in Fig.~\ref{fig:fits}.

\begin{figure}
	\centering
 \includegraphics[width=85mm, angle=0]{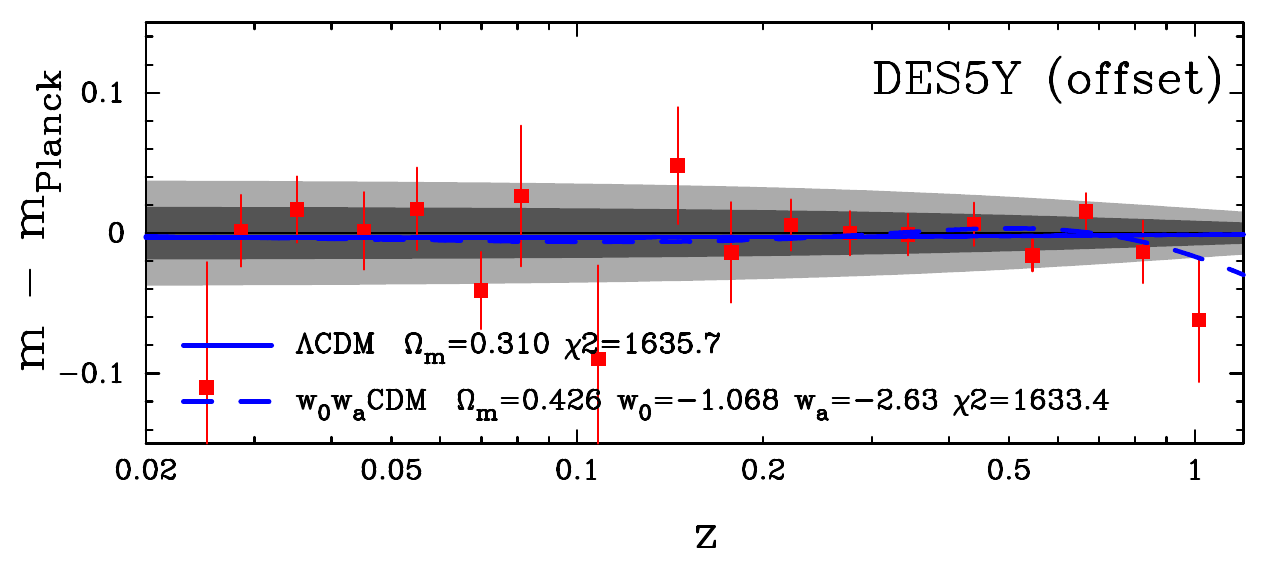} 
 \caption{Magnitude redshift residuals as in Fig.~\ref{fig:fits} but subtracting 0.04 mag. from  the low redshift subsamples
   in the DES5Y catalogue.  The best fit \LCDM\ cosmology is now indistinguishable from
   the best fit \Planck\ \LCDM\ cosmology and there is no preference for evolving dark energy.}

	  \label{fig:ODESfits}

\end{figure}

To illustrate this point, we have subtracted 0.04 mag. from the low redshift sub-samples in the the DES5Y catalogue.
The residuals in the magnitude-redshift relation after applying this offset
are plotted in Fig.~\ref{fig:ODESfits}.  This figure can be compared directly to the DES5Y plot shown on Fig.~\ref{fig:fits}.
The best fit \LCDM\ cosmology now has a value of $\Omega_m$ in near perfect agreement with the \Planck\ \LCDM\ cosmology.
Allowing $w_0$ and $w_a$ leads to a small reduction in $\chi^2$
of 2.5 (driven mainly by the low point at $z\sim 1$).
Evidently,  the recalibrated DES5Y catalogue does not show a 
statistically significant preference for a time varying equation-of-state.

\begin{figure*}
  \center
  \includegraphics[width=55mm, angle=0]{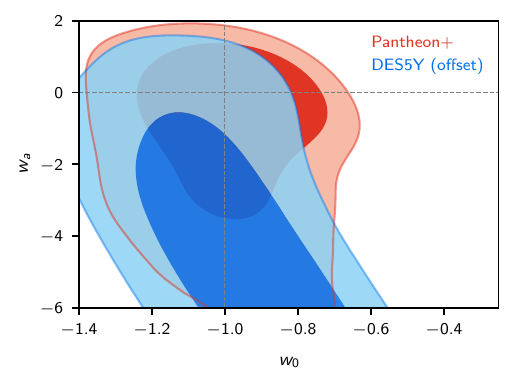}
  \includegraphics[width=55mm, angle=0]{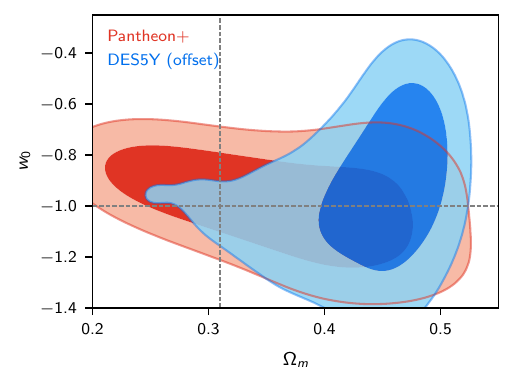}
  \includegraphics[width=55mm, angle=0]{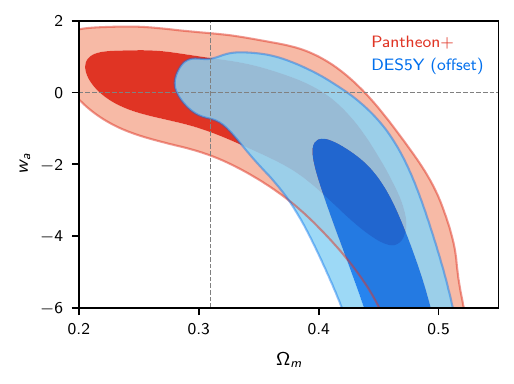}   
  \caption{Parameter constraints as in Fig.~\ref{fig:PanDES} but now using the  DES5Y sample with  $0.04$ mag. subtracted from the low redshift SN.}

	\label{fig:PanCDES}

\end{figure*}

The parameter plots of Fig.~\ref{fig:PanDES} illustrate the tension between the Pantheon+ and DES5Y SN samples.
Using the re-calibrated DES5Y catalogue we find the parameter constraints shown in Fig.~\ref{fig:PanCDES}. As expected,
there is now much greater overlap between Pantheon+ and DES5Y and both surveys are consistent with the \LCDM\ expectation of $w_0=-1$, $w_a = 0$. The corrected DESY5 constraints are still skewed to high values of $\Omega_m$ as a consequence of the negative residual  at $z > 1$
in Fig. \ref{fig:ODESfits} (which is sensitive to the large bias corrections applied at these redshifts, see Fig.~\ref{fig:bias}).

In summary, the   results presented in this Section show that reliable results on evolving dark energy require consistency in the photometry of  low and high redshift SN to an an accuracy  substantially better than $0.04$ magnitudes. This conclusion should be uncontroversial. Whether the magnitudes offsets  evident in Figs.~\ref{fig:PanDESmags} and \ref{fig:PanDES_offset} are   indicative of systematic errors in one
or both SN compilations is a more controversial question.

The sensitivity of the evidence for dynamical dark energy to 
a possible systematic photometric mismatch between the low and high redshift SN  in the DES5Y compilation has been noticed by other authors \citep{Gialamas:2024, Notari:2024}.

\section{Analysis by the DES Collaboration}
\label{sec:DESreply}

\begin{figure}
	\centering
 \includegraphics[width=85mm, angle=0]{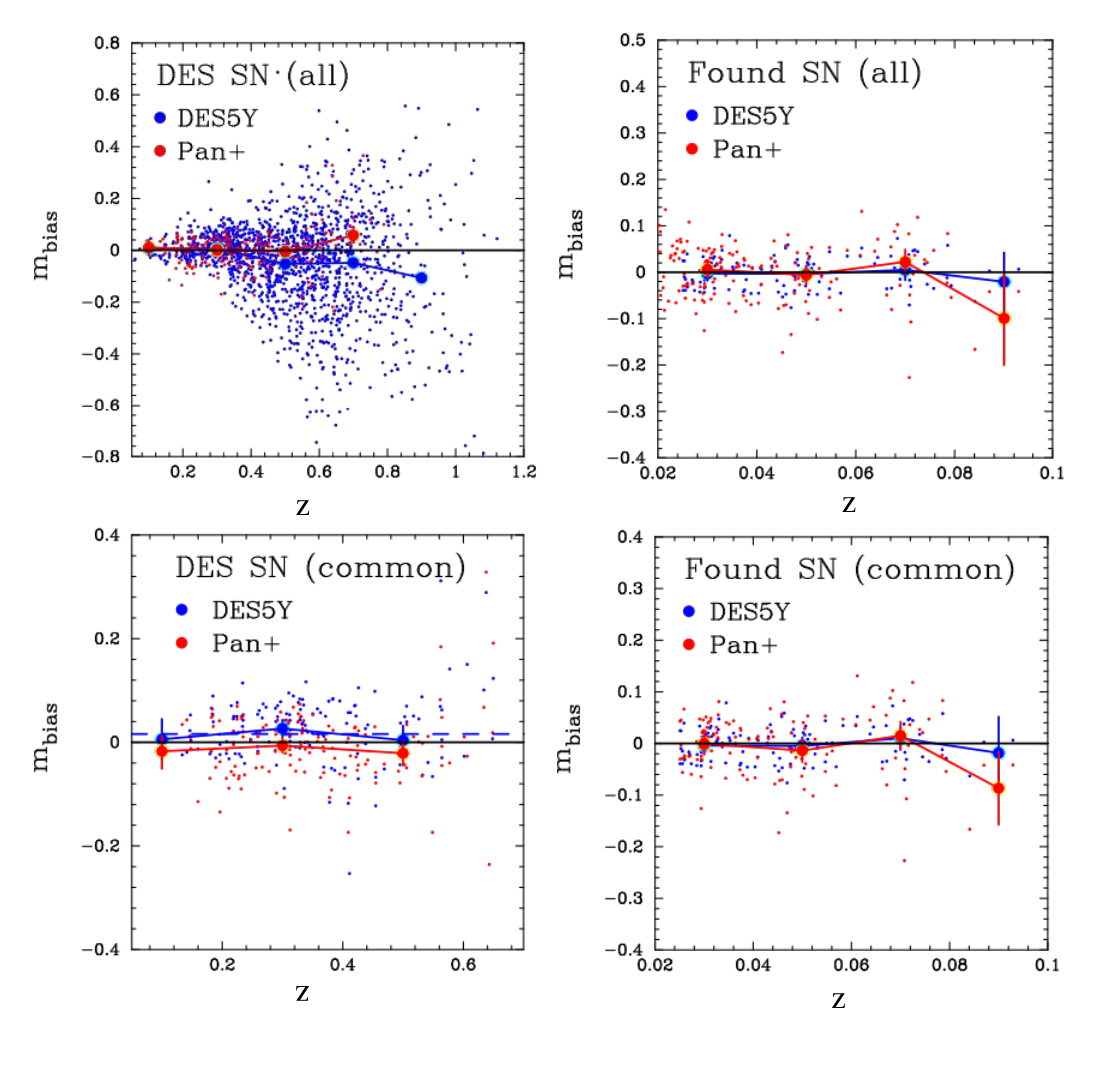} 
 \caption{The upper panels show the bias corrections for all DES (left) and Foundation (right) SN listed in the DES5Y 
 and Pantheon+ catalogues\protect\footnotemark . The small points show the individual bias
 corrections, while the large points show the maximum likelihood  bias corrections in broad redshift bins. The lower panels show the bias corrections for SN common to both samples. The dasked line in the lower left  panel shows an offset of $m_{\rm bias} = 0.016$ consistent with the differential selection function offset determined by V25.}

	  \label{fig:bias}

\end{figure}

Following the first version of this paper, the DES collaboration has carried out a detailed inventory of the 
causes of the $\sim 0.04$ magnitude shift identified in the previous section \citep[][hereafter V25]{DESreply:2025}.

V25 estimate a relative distance modulus offset of    $\Delta \mu_{\rm offset} = -0.042$ mag between DES SN common to DES5Y and Pantheon+ and low redshift SN common to both catalogues, consistent with the offset identified in the previous section.  They note that there
have been analysis improvements since the Pantheon+ catalogue was produced.  To assess their impact,  they have reanalysed the DES5Y sample using the Pantheon+ analysis choices. The inventory of the contributions to $\Delta \mu_{\rm offset}$ identified by the DES team
is summarized in their Table 1.

V25 conclude that  a contribution of $-0.016$ mag. comes from differences in the selection functions of the DES spectroscopic SN sample compared to the full DES SN  sample used in
the DES5Y compilation. Figure ~\ref{fig:bias} shows that this
number is consistent with the bias corrections  (computed from the Beams with Bias Correction, BBC,  simulations \citep{Kessler:2017}) listed in the Pantheon+ and DES5Y  catalogues.  Of the remaining offset of  $-0.026$ mag.,   V25 conclude that $-0.008$ mag.
comes from improvements to the intrinsic scatter model and  $-0.010$  mag. comes
from improvements to the host stellar mass estimates affecting mainly the Foundation SN hosts. This leaves $-0.008$ mag. unexplained. In addition, the Appendix of V25 finds a differential change of $\sim 0.01$ mag between low and high redshifts in the full Pantheon+ sample caused by updates to the BBC code introduced since the Pantheon+ 
sample was constructed.

It is unsurprising that a significant part of the offset comes from the intrinsic scatter model and the model of a `mass step' (the offset of $\sim 0.1$ magnitudes at optical wavelengths 
between SN in galaxies with stellar masses  $\simlt 10^{10} M_\odot$
and SN in galaxies with higher stellar masses) since these effects are 
inter-related and are poorly understood. DES5Y uses an updated version
\citep{Popovic:2023} of the dust based intrinsic scatter model introduced by \cite{Brout:2019}.  The changes associated with host galaxy mass estimates found by V25 are not included in the systematic error budgets of either SN compilation.

V25 find no reason to alter the DES5Y analysis. However,  their analysis confirms that there are systematic differences between
SN common to the Pantheon+ and DES5Y catalogues of a size that can have a significant effect on cosmology. The impact of these analysis differences on the full Pantheon+ catalogue has not yet been quantified.

\begin{table}

\begin{center}

  \caption{The first column lists the name of the SN sample following the nomenclature in
    \protect\cite{Scolnic:2022}. The second column lists the number of SN common to both the Pantheon+ and
    DES5Y catalogues. The third column gives the median redshift of the SN sample. Column 4 list the mean and $1\sigma$ error on 
    magnitude difference of SN in common to  Pantheon+ and DES5Y.The CFA4P3 sample has only 3 SN so we list only the mean. }

\label{tab:magfits}

\smallskip

\begin{tabular}{l|c|c|c|} \hline 
  sample  &  $N$ & $z_{\rm median}$  &  $\langle m_{\rm Pan} - m_{\rm DES}\rangle$ \\\hline
 DES5Y  & $145$ & $0.329$ & $-0.0122 \pm 0.0006$ \\
 FOUND  & $118$ & $0.043$ & $-0.0508 \pm 0.0007$ \\
 CFA3S & $14$ & $0.037$ & $-0.0344  \pm 0.0111$  \\
 CFA3K & $27$ & $0.033$ & $-0.0616  \pm 0.0163$  \\
 CFA4P2 & $18$ & $0.038$ & $-0.0547  \pm 0.0196$ \\
 CFA4P3 & $3$ &  $0.033$ & \ \ ($0.029$)   \\
 CSP & $7$ & $0.038$ & \ \ \   $0.0036  \pm 0.0207$  \\
 All low z  & $187$ & $0.038$ & $-0.0482 \pm 0.0057$ \\ \hline
\end{tabular}
\end{center}
\end{table}

\section{Discussion and Conclusions}
\label{sec:conclusions}

The elusive nature of dark energy has formed an important part of the science cases for
several ambitious projects that are either underway or will soon be taking data
\citep[e.g][]{DESI_SCIENCE:2016, Amendola:2018, Mao:2022}. The effectiveness of these
projects has often been judged in terms of a `figure of merit' defined as the inverse of the
area contained within the 95\% confidence contours in the $w_0-w_a$ plane
\citep{Albrecht:2006}. However, the figure of merit gives an incomplete picture since the area in $w_0-w_a$ space accessible to
a single, or combination,  of experiments depends on their redshift range (recognized by \cite{Albrecht:2006}
via the definition of a sample dependent pivot value of $w_p$ with which to test  for deviations from  $w=-1$, see also \cite{Cortes:2024}). In interpreting the DESI24 evidence for dynamical dark energy, it is important to consider the redshift range spanned by each dataset.

\footnotetext{Following V25 I have added $0.04$ mag. to the bias
values listed in the DES5Y catalogue to give mean biases close to zero. }

The most direct tests of evolving dark energy come from geometric
measures of the expansion history using standard rulers and standard candles.
BAO measurements apply a standard ruler to measure the comoving angular diameter distance $D_M(z)$ and Hubble parameter $H(z)$ as a function of redshift. As summarized in DESI24, BAO measurements covering the redshift range $\sim 0.3-2.3$,
including the new measurements from DESI, are consistent with a \LCDM\ expansion
history. On their own, BAO measurements do not provide any evidence for evolving dark energy. DESI BAO measurements  combined with Pantheon+ SN are consistent with \LCDM. It is only when BAO are combined with Union3 or DES5Y SN data that
a substantial preference for dynamical dark energy emerges (see Fig. 6 of DESI24 and Appendix \ref{sec:appendix}). The dynamical dark energy models preferred by these data combinations are therefore extremely sensitive to the fidelity of the SN samples.

In this paper, we have investigated  the sensitivity of 
the inferences on dark energy to  the photometric accuracy of
the SN samples.  Subtracting $0.04$ mag. from the SN
at  $z< 0.1$ (mostly consisting of the Foundation SN)  brings the magnitude-redshift relation of the DES5Y sample into good agreement with the \Planck\ \LCDM\ cosmology, eliminating any need to invoke dynamical dark energy. This part of the paper is straightforward  since it shows that systematic errors in matching the
photometry of low and high redshift SN samples must be 
substantially smaller than $0.04$ mag. if one is to claim evidence
for evolving dark energy. 

The more contentious  part of this paper involves the interpretation of differences in the  photometry of SN common to Pantheon+ and DES5Y. Such a comparison  provides an important  test for systematic errors at the $0.01$ mag. level. 
The results summarized in Fig.~\ref{fig:PanDES_offset}  do
indeed show an offset of $\sim 0.04$ magnitude between high and low redshift SN. In the original version of this paper this observation provided motivation to explore whether an offset of this size could reconcile the DES5Y sample with the \LCDM\ cosmology.

However, a simulation based analysis by the DES collaboration finds that the offset of Fig.~\ref{fig:PanDES_offset}  is caused partly by differences in the selection functions of DES SN and partly by
improvements made in DES5Y to the intrinsic scatter
model and host galaxy stellar mass estimates (not included in the systematic error budgets). V25 
therefore conclude that a significant fraction of the offset
apparent  in Fig.~\ref{fig:PanDES_offset} reflects systematic biases
in SN magnitudes listed in  the Pantheon+ catalogue.  The impact of the analysis changes on the full   Pantheon+ compilation, which has been used extensively to constrain cosmology, has not yet been assessed.

This explanation  is not entirely satisfactory. It 
ascribes  significant differences between the two catalogues to 
two aspects of the SN modelling that are 
not well understood. In particular, there is considerable evidence that the step function representation of 
the dependence on host galaxy parameters is an oversimplification
and that SN colour and luminosity are more tightly correlated with local properties such as stellar age, metallicty and  star formation rate 
\citep[see e.g.][and references therein]{Rigault:2013, Rigault:2020, Kelsey:2023}. Furthermore, these model assumptions in Pantheon+ and DES5Y
fold into the bias corrections computed from the BBC simulations, which are extremely large for DES photometrically selected SN  (see Fig.~ \ref{fig:bias}). These bias corrections must, on average, be accurate at the 1\% level if  cosmology results are to be 
unbiased. The inventory of systematics present by V25 is constructed entirely within the BBC framework, but no independent test of the accuracy of the BBC methodology and associated bias corrections, which are large for the DES photometrically classified SN,  has been presented.

Fortunately, the large complete sample of SN at $z < 0.06$ identified with  the Zwicky Transient Facility (ZTF) \citep[see the overview by][and references therein]{Rigault:2024} should lead to a much improved characterization of the intrinsic properties of Type 1a SN and their dependence on environment. Ideally, ZTF SN should be
matched with a high quality sample of more distant SN with less scatter and smaller selection biases compared to  the photometrically selected DES SN. Clearly stringent control of systematic errors is required if the SN data are to be used to claim evidence for new physics.

\section{Acknowledgements}
I thank the Leverhulme Foundation for the award of a Leverhulme Emeritus Fellowship.
I am especially grateful to Maria Vincenzi,  Paul Shah and Dan Scolnic for answering questions on the DES5Y analysis. I thank
Sesh Nadathur for comments on the DESI comparison reported  in Appendix \ref{sec:appendix}.  I thank Lisa Kelsey, Mickael Rigault  and Suhail Dhawan for discussions concerning systematic errors in SN surveys.

\section*{Data Availability} 

No new data were generated or analysed in support of this research.

\begin{figure*}
  \centering
  \includegraphics[width=55mm, angle=0]{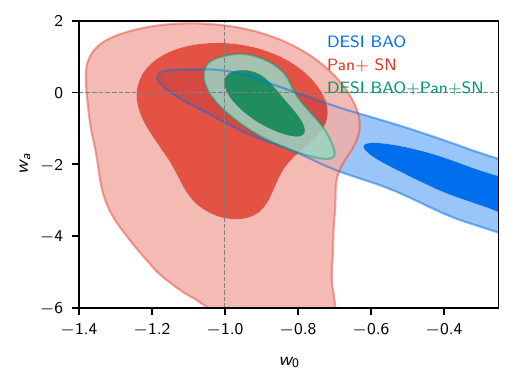}
  \includegraphics[width=55mm, angle=0]{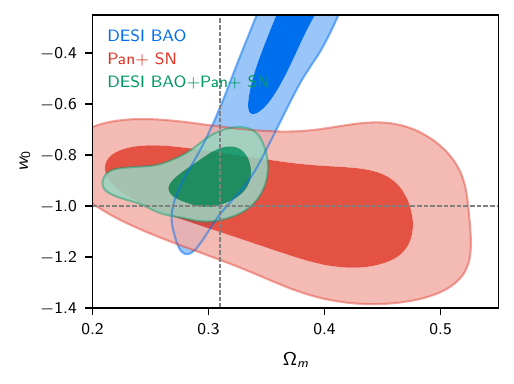}
  \includegraphics[width=55mm, angle=0]{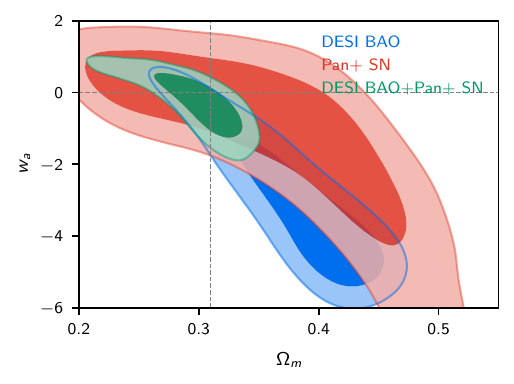}
\\  
  \includegraphics[width=55mm, angle=0]{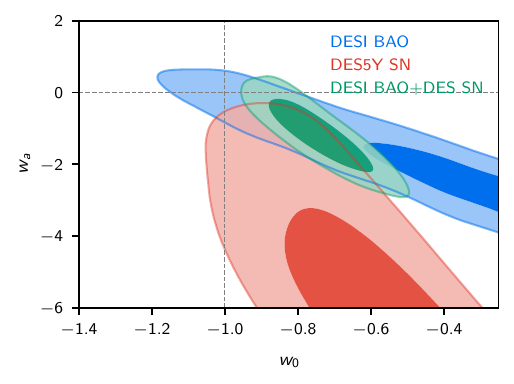}
  \includegraphics[width=55mm, angle=0]{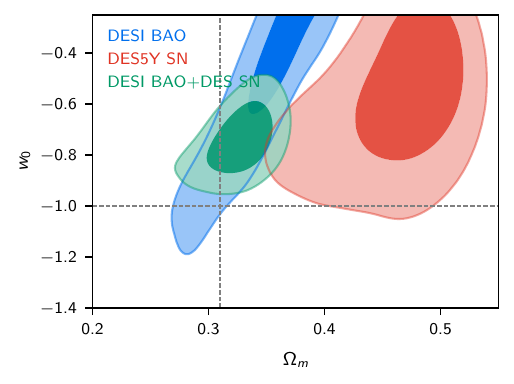}
  \includegraphics[width=55mm, angle=0]{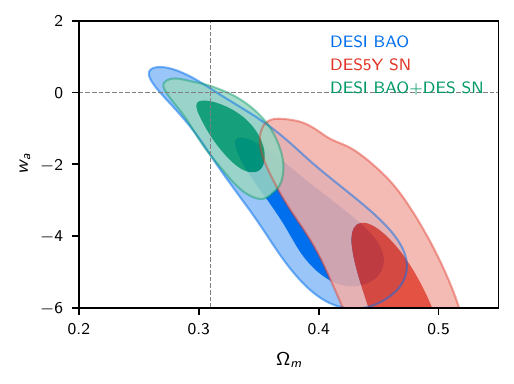}    \\
  
  \caption{68\% and 95\% contours for the marginalized posterior parameters in the $w_0-w_a$, $w_0-\Omega_m$ and
    $w_a-\Omega_m$ planes for the DESI BAO measurements
    of DESI24 (blue contours) and  Pantheon+ SN  (upper plots) and DES5Y  SN (lower plots). The green contours show the joint BAO+SN
    constraints. The dashed lines show the values for
  the best fit \Planck\ \LCDM\ cosmology.}

	\label{fig:PanDESDESI}

\end{figure*}

\bibliographystyle{mnras}
\bibliography{SNpaper}

\begin{thebibliography}{}
\makeatletter
\relax
\def\mn@urlcharsother{\let\do\@makeother \do\$\do\&\do\#\do\^\do\_\do\%\do\~}
\def\mn@doi{\begingroup\mn@urlcharsother \@ifnextchar [ {\mn@doi@}
  {\mn@doi@[]}}
\def\mn@doi@[#1]#2{\def\@tempa{#1}\ifx\@tempa\@empty \href
  {http://dx.doi.org/#2} {doi:#2}\else \href {http://dx.doi.org/#2} {#1}\fi
  \endgroup}
\def\mn@eprint#1#2{\mn@eprint@#1:#2::\@nil}
\def\mn@eprint@arXiv#1{\href {http://arxiv.org/abs/#1} {{\tt arXiv:#1}}}
\def\mn@eprint@dblp#1{\href {http://dblp.uni-trier.de/rec/bibtex/#1.xml}
  {dblp:#1}}
\def\mn@eprint@#1:#2:#3:#4\@nil{\def\@tempa {#1}\def\@tempb {#2}\def\@tempc
  {#3}\ifx \@tempc \@empty \let \@tempc \@tempb \let \@tempb \@tempa \fi \ifx
  \@tempb \@empty \def\@tempb {arXiv}\fi \@ifundefined
  {mn@eprint@\@tempb}{\@tempb:\@tempc}{\expandafter \expandafter \csname
  mn@eprint@\@tempb\endcsname \expandafter{\@tempc}}}

\bibitem[\protect\citeauthoryear{{Alam} et~al.,}{{Alam}
  et~al.}{2017}]{Alam:2017}
{Alam} S.,  et~al., 2017, \mn@doi [\mnras] {10.1093/mnras/stx721}, \href
  {https://ui.adsabs.harvard.edu/abs/2017MNRAS.470.2617A} {470, 2617}

\bibitem[\protect\citeauthoryear{{Alam} et~al.,}{{Alam}
  et~al.}{2021}]{Alam:2021}
{Alam} S.,  et~al., 2021, \mn@doi [\prd] {10.1103/PhysRevD.103.083533}, \href
  {https://ui.adsabs.harvard.edu/abs/2021PhRvD.103h3533A} {103, 083533}

\bibitem[\protect\citeauthoryear{{Albrecht} et~al.,}{{Albrecht}
  et~al.}{2006}]{Albrecht:2006}
{Albrecht} A.,  et~al., 2006, \mn@doi [arXiv e-prints]
  {10.48550/arXiv.astro-ph/0609591}, \href
  {https://ui.adsabs.harvard.edu/abs/2006astro.ph..9591A} {pp
  astro--ph/0609591}

\bibitem[\protect\citeauthoryear{{Amendola} et~al.,}{{Amendola}
  et~al.}{2018}]{Amendola:2018}
{Amendola} L.,  et~al., 2018, \mn@doi [Living Reviews in Relativity]
  {10.1007/s41114-017-0010-3}, \href
  {https://ui.adsabs.harvard.edu/abs/2018LRR....21....2A} {21, 2}

\bibitem[\protect\citeauthoryear{{Betoule} et~al.,}{{Betoule}
  et~al.}{2014}]{Betoule:2014}
{Betoule} M.,  et~al., 2014, \mn@doi [\aap] {10.1051/0004-6361/201423413},
  \href {https://ui.adsabs.harvard.edu/abs/2014A&A...568A..22B} {568, A22}

\bibitem[\protect\citeauthoryear{{Brout} \& {Scolnic}}{{Brout} \&
  {Scolnic}}{2021}]{Brout:2021}
{Brout} D.,  {Scolnic} D.,  2021, \mn@doi [\apj] {10.3847/1538-4357/abd69b},
  \href {https://ui.adsabs.harvard.edu/abs/2021ApJ...909...26B} {909, 26}

\bibitem[\protect\citeauthoryear{{Brout} et~al.,}{{Brout}
  et~al.}{2019}]{Brout:2019}
{Brout} D.,  et~al., 2019, \mn@doi [\apj] {10.3847/1538-4357/ab08a0}, \href
  {https://ui.adsabs.harvard.edu/abs/2019ApJ...874..150B} {874, 150}

\bibitem[\protect\citeauthoryear{{Brout} et~al.,}{{Brout}
  et~al.}{2022a}]{Brout:2022}
{Brout} D.,  et~al., 2022a, \mn@doi [\apj] {10.3847/1538-4357/ac8e04}, \href
  {https://ui.adsabs.harvard.edu/abs/2022ApJ...938..110B} {938, 110}

\bibitem[\protect\citeauthoryear{{Brout} et~al.,}{{Brout}
  et~al.}{2022b}]{Brout:2022b}
{Brout} D.,  et~al., 2022b, \mn@doi [\apj] {10.3847/1538-4357/ac8bcc}, \href
  {https://ui.adsabs.harvard.edu/abs/2022ApJ...938..111B} {938, 111}

\bibitem[\protect\citeauthoryear{{Chevallier} \& {Polarski}}{{Chevallier} \&
  {Polarski}}{2001}]{Chevallier:2001}
{Chevallier} M.,  {Polarski} D.,  2001, \mn@doi [International Journal of
  Modern Physics D] {10.1142/S0218271801000822}, \href
  {https://ui.adsabs.harvard.edu/abs/2001IJMPD..10..213C} {10, 213}

\bibitem[\protect\citeauthoryear{{Conley} et~al.,}{{Conley}
  et~al.}{2011}]{Conley:2011}
{Conley} A.,  et~al., 2011, \mn@doi [\apjs] {10.1088/0067-0049/192/1/1}, \href
  {https://ui.adsabs.harvard.edu/abs/2011ApJS..192....1C} {192, 1}

\bibitem[\protect\citeauthoryear{{Copeland}, {Sami}  \& {Tsujikawa}}{{Copeland}
  et~al.}{2006}]{Copeland:2006}
{Copeland} E.~J.,  {Sami} M.,   {Tsujikawa} S.,  2006, \mn@doi [International
  Journal of Modern Physics D] {10.1142/S021827180600942X}, \href
  {https://ui.adsabs.harvard.edu/abs/2006IJMPD..15.1753C} {15, 1753}

\bibitem[\protect\citeauthoryear{{Cort{\^e}s} \& {Liddle}}{{Cort{\^e}s} \&
  {Liddle}}{2024}]{Cortes:2024}
{Cort{\^e}s} M.,  {Liddle} A.~R.,  2024, \mn@doi [arXiv e-prints]
  {10.48550/arXiv.2404.08056}, \href
  {https://ui.adsabs.harvard.edu/abs/2024arXiv240408056C} {p. arXiv:2404.08056}

\bibitem[\protect\citeauthoryear{{D'Amico}, {Donath}, {Lewandowski}, {Senatore}
   \& {Zhang}}{{D'Amico} et~al.}{2024}]{dAmico:2024}
{D'Amico} G.,  {Donath} Y.,  {Lewandowski} M.,  {Senatore} L.,   {Zhang} P.,
  2024, \mn@doi [\jcap] {10.1088/1475-7516/2024/05/059}, \href
  {https://ui.adsabs.harvard.edu/abs/2024JCAP...05..059D} {2024, 059}

\bibitem[\protect\citeauthoryear{{DES Collaboration} et~al.,}{{DES
  Collaboration} et~al.}{2024}]{DES5Y:2024}
{DES Collaboration} et~al., 2024, \mn@doi [arXiv e-prints]
  {10.48550/arXiv.2401.02929}, \href
  {https://ui.adsabs.harvard.edu/abs/2024arXiv240102929D} {p. arXiv:2401.02929}

\bibitem[\protect\citeauthoryear{{DESI Collaboration} et~al.,}{{DESI
  Collaboration} et~al.}{2016}]{DESI_SCIENCE:2016}
{DESI Collaboration} et~al., 2016, \mn@doi [arXiv e-prints]
  {10.48550/arXiv.1611.00036}, \href
  {https://ui.adsabs.harvard.edu/abs/2016arXiv161100036D} {p. arXiv:1611.00036}

\bibitem[\protect\citeauthoryear{{DESI Collaboration} et~al.,}{{DESI
  Collaboration} et~al.}{2024a}]{DESI_1:2024}
{DESI Collaboration} et~al., 2024a, \mn@doi [arXiv e-prints]
  {10.48550/arXiv.2404.03000}, \href
  {https://ui.adsabs.harvard.edu/abs/2024arXiv240403000D} {p. arXiv:2404.03000}

\bibitem[\protect\citeauthoryear{{DESI Collaboration} et~al.,}{{DESI
  Collaboration} et~al.}{2024b}]{DESI_2:2024}
{DESI Collaboration} et~al., 2024b, \mn@doi [arXiv e-prints]
  {10.48550/arXiv.2404.03001}, \href
  {https://ui.adsabs.harvard.edu/abs/2024arXiv240403001D} {p. arXiv:2404.03001}

\bibitem[\protect\citeauthoryear{{DESI Collaboration} et~al.,}{{DESI
  Collaboration} et~al.}{2024c}]{DESI:2024}
{DESI Collaboration} et~al., 2024c, \mn@doi [arXiv e-prints]
  {10.48550/arXiv.2404.03002}, \href
  {https://ui.adsabs.harvard.edu/abs/2024arXiv240403002D} {p. arXiv:2404.03002}

\bibitem[\protect\citeauthoryear{{Dhawan}, {Popovic}  \& {Goobar}}{{Dhawan}
  et~al.}{2024}]{Dhawan:2024}
{Dhawan} S.,  {Popovic} B.,   {Goobar} A.,  2024, \mn@doi [arXiv e-prints]
  {10.48550/arXiv.2409.18668}, \href
  {https://ui.adsabs.harvard.edu/abs/2024arXiv240918668D} {p. arXiv:2409.18668}

\bibitem[\protect\citeauthoryear{{Dyson}, {Kleban}  \& {Susskind}}{{Dyson}
  et~al.}{2002}]{Dyson:2002}
{Dyson} L.,  {Kleban} M.,   {Susskind} L.,  2002, \mn@doi [Journal of High
  Energy Physics] {10.1088/1126-6708/2002/10/011}, \href
  {https://ui.adsabs.harvard.edu/abs/2002JHEP...10..011D} {2002, 011}

\bibitem[\protect\citeauthoryear{{Efstathiou}}{{Efstathiou}}{2024}]{Efstathiou:2024}
{Efstathiou} G.,  2024, \mn@doi [arXiv e-prints] {10.48550/arXiv.2406.12106},
  \href {https://ui.adsabs.harvard.edu/abs/2024arXiv240612106E} {p.
  arXiv:2406.12106}

\bibitem[\protect\citeauthoryear{{Efstathiou} \& {Gratton}}{{Efstathiou} \&
  {Gratton}}{2021}]{Efstathiou:2021}
{Efstathiou} G.,  {Gratton} S.,  2021, \mn@doi [The Open Journal of
  Astrophysics] {10.21105/astro.1910.00483}, \href
  {https://ui.adsabs.harvard.edu/abs/2021OJAp....4E...8E} {4, 8}

\bibitem[\protect\citeauthoryear{{Feroz}, {Hobson}  \& {Bridges}}{{Feroz}
  et~al.}{2009}]{Feroz:2009}
{Feroz} F.,  {Hobson} M.~P.,   {Bridges} M.,  2009, \mn@doi [\mnras]
  {10.1111/j.1365-2966.2009.14548.x}, \href
  {https://ui.adsabs.harvard.edu/abs/2009MNRAS.398.1601F} {398, 1601}

\bibitem[\protect\citeauthoryear{{Feroz}, {Hobson}  \& {Bridges}}{{Feroz}
  et~al.}{2011}]{Feroz:2011}
{Feroz} F.,  {Hobson} M.~P.,   {Bridges} M.,  2011, {MultiNest: Efficient and
  Robust Bayesian Inference} (\mn@eprint {ascl} {1109.006})

\bibitem[\protect\citeauthoryear{{Foley} et~al.,}{{Foley}
  et~al.}{2018}]{Foley:2018}
{Foley} R.~J.,  et~al., 2018, \mn@doi [\mnras] {10.1093/mnras/stx3136}, \href
  {https://ui.adsabs.harvard.edu/abs/2018MNRAS.475..193F} {475, 193}

\bibitem[\protect\citeauthoryear{{Foreman-Mackey} et~al.,}{{Foreman-Mackey}
  et~al.}{2013}]{emcee:2013}
{Foreman-Mackey} D.,  et~al., 2013, {emcee: The MCMC Hammer}, Astrophysics
  Source Code Library, record ascl:1303.002

\bibitem[\protect\citeauthoryear{{Gialamas}, {H{\"u}tsi}, {Kannike},
  {Racioppi}, {Raidal}, {Vasar}  \& {Veerm{\"a}e}}{{Gialamas}
  et~al.}{2024}]{Gialamas:2024}
{Gialamas} I.~D.,  {H{\"u}tsi} G.,  {Kannike} K.,  {Racioppi} A.,  {Raidal} M.,
   {Vasar} M.,   {Veerm{\"a}e} H.,  2024, \mn@doi [arXiv e-prints]
  {10.48550/arXiv.2406.07533}, \href
  {https://ui.adsabs.harvard.edu/abs/2024arXiv240607533G} {p. arXiv:2406.07533}

\bibitem[\protect\citeauthoryear{{Guy} et~al.,}{{Guy} et~al.}{2007}]{Guy:2007}
{Guy} J.,  et~al., 2007, \mn@doi [\aap] {10.1051/0004-6361:20066930}, \href
  {https://ui.adsabs.harvard.edu/abs/2007A&A...466...11G} {466, 11}

\bibitem[\protect\citeauthoryear{{Hicken} et~al.,}{{Hicken}
  et~al.}{2009}]{Hicken:2009}
{Hicken} M.,  et~al., 2009, \mn@doi [\apj] {10.1088/0004-637X/700/1/331}, \href
  {https://ui.adsabs.harvard.edu/abs/2009ApJ...700..331H} {700, 331}

\bibitem[\protect\citeauthoryear{{Hicken} et~al.,}{{Hicken}
  et~al.}{2012}]{Hicken:2012}
{Hicken} M.,  et~al., 2012, \mn@doi [\apjs] {10.1088/0067-0049/200/2/12}, \href
  {https://ui.adsabs.harvard.edu/abs/2012ApJS..200...12H} {200, 12}

\bibitem[\protect\citeauthoryear{{Kelsey} et~al.,}{{Kelsey}
  et~al.}{2023}]{Kelsey:2023}
{Kelsey} L.,  et~al., 2023, \mn@doi [\mnras] {10.1093/mnras/stac3711}, \href
  {https://ui.adsabs.harvard.edu/abs/2023MNRAS.519.3046K} {519, 3046}

\bibitem[\protect\citeauthoryear{{Kenworthy} et~al.,}{{Kenworthy}
  et~al.}{2021}]{Kenworthy:2021}
{Kenworthy} W.~D.,  et~al., 2021, \mn@doi [\apj] {10.3847/1538-4357/ac30d8},
  \href {https://ui.adsabs.harvard.edu/abs/2021ApJ...923..265K} {923, 265}

\bibitem[\protect\citeauthoryear{{Kessler} \& {Scolnic}}{{Kessler} \&
  {Scolnic}}{2017}]{Kessler:2017}
{Kessler} R.,  {Scolnic} D.,  2017, \mn@doi [\apj]
  {10.3847/1538-4357/836/1/56}, \href
  {https://ui.adsabs.harvard.edu/abs/2017ApJ...836...56K} {836, 56}

\bibitem[\protect\citeauthoryear{{Krisciunas} et~al.,}{{Krisciunas}
  et~al.}{2017}]{Krisciunas:2017}
{Krisciunas} K.,  et~al., 2017, \mn@doi [\aj] {10.3847/1538-3881/aa8df0}, \href
  {https://ui.adsabs.harvard.edu/abs/2017AJ....154..211K} {154, 211}

\bibitem[\protect\citeauthoryear{{Linder}}{{Linder}}{2003}]{Linder:2003}
{Linder} E.~V.,  2003, \mn@doi [\prl] {10.1103/PhysRevLett.90.091301}, \href
  {https://ui.adsabs.harvard.edu/abs/2003PhRvL..90i1301L} {90, 091301}

\bibitem[\protect\citeauthoryear{{Mantz} et~al.,}{{Mantz}
  et~al.}{2022}]{Mantz:2022}
{Mantz} A.~B.,  et~al., 2022, \mn@doi [\mnras] {10.1093/mnras/stab3390}, \href
  {https://ui.adsabs.harvard.edu/abs/2022MNRAS.510..131M} {510, 131}

\bibitem[\protect\citeauthoryear{Mao et~al.,}{Mao et~al.}{2022}]{Mao:2022}
Mao Y.-Y.,  et~al., 2022, Snowmass2021: Vera C. Rubin Observatory as a Flagship
  Dark Matter Experiment (\mn@eprint {arXiv} {2203.07252}), \url
  {https://arxiv.org/abs/2203.07252}

\bibitem[\protect\citeauthoryear{{M{\"o}ller} \& {de
  Boissi{\`e}re}}{{M{\"o}ller} \& {de Boissi{\`e}re}}{2020}]{Moller:2020}
{M{\"o}ller} A.,  {de Boissi{\`e}re} T.,  2020, \mn@doi [\mnras]
  {10.1093/mnras/stz3312}, \href
  {https://ui.adsabs.harvard.edu/abs/2020MNRAS.491.4277M} {491, 4277}

\bibitem[\protect\citeauthoryear{{Notari}, {Redi}  \& {Tesi}}{{Notari}
  et~al.}{2024}]{Notari:2024}
{Notari} A.,  {Redi} M.,   {Tesi} A.,  2024, \mn@doi [arXiv e-prints]
  {10.48550/arXiv.2411.11685}, \href
  {https://ui.adsabs.harvard.edu/abs/2024arXiv241111685N} {p. arXiv:2411.11685}

\bibitem[\protect\citeauthoryear{{Perlmutter} et~al.,}{{Perlmutter}
  et~al.}{1999}]{Perlmutter:1999}
{Perlmutter} S.,  et~al., 1999, \mn@doi [\apj] {10.1086/307221}, \href
  {https://ui.adsabs.harvard.edu/abs/1999ApJ...517..565P} {517, 565}

\bibitem[\protect\citeauthoryear{{Philcox} \& {Ivanov}}{{Philcox} \&
  {Ivanov}}{2022}]{Philcox:2022}
{Philcox} O. H.~E.,  {Ivanov} M.~M.,  2022, \mn@doi [\prd]
  {10.1103/PhysRevD.105.043517}, \href
  {https://ui.adsabs.harvard.edu/abs/2022PhRvD.105d3517P} {105, 043517}

\bibitem[\protect\citeauthoryear{{Planck Collaboration} et~al.,}{{Planck
  Collaboration} et~al.}{2020}]{Planck_Params_2018}
{Planck Collaboration} et~al., 2020, \mn@doi [\aap]
  {10.1051/0004-6361/201833910}, \href
  {https://ui.adsabs.harvard.edu/abs/2020A&A...641A...6P} {641, A6}

\bibitem[\protect\citeauthoryear{{Popovic}, {Brout}, {Kessler}, {Scolnic}  \&
  {Lu}}{{Popovic} et~al.}{2021}]{Popovic:2021}
{Popovic} B.,  {Brout} D.,  {Kessler} R.,  {Scolnic} D.,   {Lu} L.,  2021,
  \mn@doi [\apj] {10.3847/1538-4357/abf14f}, \href
  {https://ui.adsabs.harvard.edu/abs/2021ApJ...913...49P} {913, 49}

\bibitem[\protect\citeauthoryear{{Popovic}, {Brout}, {Kessler}  \&
  {Scolnic}}{{Popovic} et~al.}{2023}]{Popovic:2023}
{Popovic} B.,  {Brout} D.,  {Kessler} R.,   {Scolnic} D.,  2023, \mn@doi [\apj]
  {10.3847/1538-4357/aca273}, \href
  {https://ui.adsabs.harvard.edu/abs/2023ApJ...945...84P} {945, 84}

\bibitem[\protect\citeauthoryear{{Riess} et~al.,}{{Riess}
  et~al.}{1998}]{Riess:1998}
{Riess} A.~G.,  et~al., 1998, \mn@doi [\aj] {10.1086/300499}, \href
  {https://ui.adsabs.harvard.edu/abs/1998AJ....116.1009R} {116, 1009}

\bibitem[\protect\citeauthoryear{{Riess} et~al.,}{{Riess}
  et~al.}{2022}]{Riess:2022}
{Riess} A.~G.,  et~al., 2022, \mn@doi [\apjl] {10.3847/2041-8213/ac5c5b}, \href
  {https://ui.adsabs.harvard.edu/abs/2022ApJ...934L...7R} {934, L7}

\bibitem[\protect\citeauthoryear{{Rigault} et~al.,}{{Rigault}
  et~al.}{2013}]{Rigault:2013}
{Rigault} M.,  et~al., 2013, \mn@doi [\aap] {10.1051/0004-6361/201322104},
  \href {https://ui.adsabs.harvard.edu/abs/2013A&A...560A..66R} {560, A66}

\bibitem[\protect\citeauthoryear{{Rigault} et~al.,}{{Rigault}
  et~al.}{2020}]{Rigault:2020}
{Rigault} M.,  et~al., 2020, \mn@doi [\aap] {10.1051/0004-6361/201730404},
  \href {https://ui.adsabs.harvard.edu/abs/2020A&A...644A.176R} {644, A176}

\bibitem[\protect\citeauthoryear{{Rigault} et~al.,}{{Rigault}
  et~al.}{2024}]{Rigault:2024}
{Rigault} M.,  et~al., 2024, \mn@doi [arXiv e-prints]
  {10.48550/arXiv.2409.04346}, \href
  {https://ui.adsabs.harvard.edu/abs/2024arXiv240904346R} {p. arXiv:2409.04346}

\bibitem[\protect\citeauthoryear{{Rubin} et~al.,}{{Rubin}
  et~al.}{2023}]{Rubin:2023}
{Rubin} D.,  et~al., 2023, \mn@doi [arXiv e-prints]
  {10.48550/arXiv.2311.12098}, \href
  {https://ui.adsabs.harvard.edu/abs/2023arXiv231112098R} {p. arXiv:2311.12098}

\bibitem[\protect\citeauthoryear{{S{\'a}nchez} et~al.,}{{S{\'a}nchez}
  et~al.}{2024}]{Sanchez:2024}
{S{\'a}nchez} B.~O.,  et~al., 2024, \mn@doi [arXiv e-prints]
  {10.48550/arXiv.2406.05046}, \href
  {https://ui.adsabs.harvard.edu/abs/2024arXiv240605046S} {p. arXiv:2406.05046}

\bibitem[\protect\citeauthoryear{{Scolnic} et~al.,}{{Scolnic}
  et~al.}{2018}]{Scolnic:2018}
{Scolnic} D.~M.,  et~al., 2018, \mn@doi [\apj] {10.3847/1538-4357/aab9bb},
  \href {https://ui.adsabs.harvard.edu/abs/2018ApJ...859..101S} {859, 101}

\bibitem[\protect\citeauthoryear{{Scolnic} et~al.,}{{Scolnic}
  et~al.}{2022}]{Scolnic:2022}
{Scolnic} D.,  et~al., 2022, \mn@doi [\apj] {10.3847/1538-4357/ac8b7a}, \href
  {https://ui.adsabs.harvard.edu/abs/2022ApJ...938..113S} {938, 113}

\bibitem[\protect\citeauthoryear{{Smith} et~al.,}{{Smith}
  et~al.}{2020}]{Smith:2020}
{Smith} M.,  et~al., 2020, \mn@doi [\aj] {10.3847/1538-3881/abc01b}, \href
  {https://ui.adsabs.harvard.edu/abs/2020AJ....160..267S} {160, 267}

\bibitem[\protect\citeauthoryear{{Suzuki} et~al.,}{{Suzuki}
  et~al.}{2012}]{Suzuki:2012}
{Suzuki} N.,  et~al., 2012, \mn@doi [\apj] {10.1088/0004-637X/746/1/85}, \href
  {https://ui.adsabs.harvard.edu/abs/2012ApJ...746...85S} {746, 85}

\bibitem[\protect\citeauthoryear{{Vincenzi} et~al.,}{{Vincenzi}
  et~al.}{2024}]{Vincenzi:2024}
{Vincenzi} M.,  et~al., 2024, \mn@doi [arXiv e-prints]
  {10.48550/arXiv.2401.02945}, \href
  {https://ui.adsabs.harvard.edu/abs/2024arXiv240102945V} {p. arXiv:2401.02945}

\bibitem[\protect\citeauthoryear{{Vincenzi} et~al.,}{{Vincenzi}
  et~al.}{2025}]{DESreply:2025}
{Vincenzi} M.,  et~al., 2025, \mn@doi [arXiv e-prints]
  {10.48550/arXiv.2501.06664}, \href
  {https://ui.adsabs.harvard.edu/abs/2025arXiv250106664V} {p. arXiv:2501.06664}

\bibitem[\protect\citeauthoryear{{Weinberg}}{{Weinberg}}{1989}]{Weinberg:1989}
{Weinberg} S.,  1989, \mn@doi [Reviews of Modern Physics]
  {10.1103/RevModPhys.61.1}, \href
  {https://ui.adsabs.harvard.edu/abs/1989RvMP...61....1W} {61, 1}

\makeatother
\end{thebibliography}

\appendix
\section{Comparison of DESI BAO with Pantheon+ and DES5Y}
\label{sec:appendix}

Fig. 1 of V25 shows shows their version of the binned Hubble residuals of the Pantheon+ and DES5Y catalogues which can be compared to Fig.~\ref{fig:fits} of this paper. They  
conclude that the residuals are consistent with each other to within 1$\sigma$ in every bin giving the impression that the two SN catalogues are statistically equivalent. However, the binned Hubble residuals shown in Fig.~\ref{fig:fits} are completely
consistent with the full likelihoods which which show that it is
the Pantheon+ catalogue is neutral towards evolving dark energy
and it is DES5Y  that is responsible for the high significance
pull towards dark energy evolution reported in DESI24. 

To emphasise this point, the red contours in Fig.~{ref:PanDESDESI}
show the constraints in the space of $w_0, w_a, \Omega_m$ from the 
Pantheon+ and DES5Y SN data as plotted in Fig.~\ref{fig:PanDESDESI}. The blue contours show the constraints derived from DESI BAO (where I have used a Gaussian \Planck\ prior on the sound horizon $r_d =147.31 \pm 0.31 \ {\rm Mpc}$). As noted in DESI24, the DESI BAO measurements, show a mild preference for evolving dark energy. However, when combined with the Pantheon+ SN (green contours in the upper panel) the
\Planck\ \LCDM\ model touches the $1\sigma$ contours in all three
projections.  In other words, the Pantheon+ SN data cut off the extreme values of $w_0, w_a and \Omega_m$ along the degeneracy directions
of the DESI BAO leading to a cosmology with parameters close to 
those of the best fit \Planck\ \LCDM\ model. In contrast, the DESY contours shown in the lower panel of Fig.~\ref{fig:PanDESDESI} are
displaced from the DESI BAO contours. This is particularly apparent in the lower middle plot showing the $w_0-\Omega_m$ projection. Note
that the high values of $\Omega_m$ favoured by the DESYSN are disfavoured by DESI BAO and by the combined constraints shown by the
green contours. This is also true for the combination of DESI BAO+\Planck\  which gives the constraint $\Omega_m = 0.0344 ^{+0.032}_{-0.027}$ for models with varying $w_0$ and $w_a$. Fig.~\ref{fig:PanDESDESI} shows clearly that the $\sim 3 \sigma$ evidence for dynamical dark energy apparent in lower panel is
driven by the DES5Y SN.

\end{document}